\documentclass[twocolumn, twocoluappendix]{aastex631}

\PassOptionsToPackage{dvipsnames}{xcolor}
\usepackage{soul}

\begin{document}

\title{TeV spectral bump of cosmic-ray protons and helium nuclei: the role of nearby supernova remnants}

\author[0000-0002-0044-9751]{Sourav Bhadra$^\bigstar$}
\affiliation{Raman Research Institute, Sadashiva Nagar, Bangalore 560080, India}
\affiliation{Department of Physics, Indian Institute of Science, Bangalore 560012, India}
\email{$^\bigstar$sbhadra07@gmail.com}

%\footnote{Corresponding authors email id: sbhadra07@gmail.com, satyendra.thoudam@ku.ac.ae}:
\author[0000-0002-7066-3614]{Satyendra Thoudam$^\#$}
\affiliation{Department of Physics, Khalifa University of Science and Technology, PO Box 127788, Abu Dhabi, United Arab Emirates.}
\email{$^\#$satyendra.thoudam@ku.ac.ae}

%\collaboration{20}{(AAS Journals Data Editors)}

\author{Biman B Nath}
\affiliation{Raman Research Institute, Sadashiva Nagar, Bangalore 560080, India}

\author{Prateek Sharma}
%\altaffiliation{AASTeX v6+ programmer}
\affiliation{Department of Physics, Indian Institute of Science, Bangalore 560012, India}



\begin{abstract}

The existence of nearby discrete cosmic-ray sources can lead to many interesting effects on the observed properties of cosmic rays. Recent measurements of cosmic rays with the CALET and the DAMPE experiments have revealed a bump-like new feature in the proton and helium spectra in the energy range of $\sim\,(1-100)$~TeV/nucleon. The origin of the feature is not clearly understood. In this paper, considering an improved and more detailed analysis than previous works, and using an updated age and distance estimates of nearby supernova remnants along with an energy-dependent escape process for  cosmic rays from the remnants, we show that the spectral bump can be explained by the contribution of cosmic rays from the nearby supernova remnants, in particular, the Vela remnant. We also show that the contribution from the nearby remnants agrees well with the observed spectra of the heavier cosmic-ray elements from carbon to iron as well as with the measured all-particle cosmic-ray spectrum beyond the knee region when combined with a background flux of cosmic rays originating from distant supernova remnants.

\end{abstract}



\section{Introduction} \label{sec:intro}

Cosmic rays (hereafter CRs) represent high-energy charged particles spanning a broad energy spectrum from $1$ GeV to $\sim 10^{11}$ GeV. Supernova remnants (hereafter SNRs) have long been considered as the most promising sources of CRs in the Galaxy, particularly up to a few PeV \citep{Lagage1983}. Based on the fundamental principles of the  diffusive shock acceleration (DSA) theory of CRs \citep{Axford1977, Blandford1978} and the nature of CR transport in the Galaxy, CRs are expected to follow a power-law spectrum. This prediction is in general good agreement with the measured CR spectrum at the Earth which has an index of $\sim -2.7$ up to approximately $3\times 10^6$~GeV ($3$~PeV), commonly referred to as the CR `knee'. The spectrum steepens to $\sim -3.1$ above the knee,  possibly due to the subsequent cut-offs in the energy spectra of the different CR elements \citep{Thoudam2016}. The spectrum then  flattens back to $\sim -2.7$ at $4\times 10^9$ GeV, a feature  known as the CR `ankle', which is possibly caused by the interaction of extra-galactic protons with the cosmic microwave background (CMB) photons producing electron-positron pairs \citep{Berezinskii1988} or due to the photo-disintegration of CRs inside compact extra-galactic sources \citep{Globus2015, Unger2015}. At $\sim 10^{11}$ GeV, the spectrum shows a steepening, the so-called `GZK cutoff', which can be the result of extra-galactic CRs interacting with the CMB producing photo-pions \citep{Greisen1966, Zatsepin1966}.

Several additional distinctive features have been  discovered in the spectra of the individual CR elements below the knee energy. In this paper, we focus on the recent observations of a bump-like feature in the spectra of protons and helium nuclei by the DAMPE \citep{An2019, Alemanno2021} and the CALET \citep{Adriani2022p, Adriani2023} experiments. This feature exhibits a hardening in the spectra at a few hundred GeV, confirming earlier detections by the PAMELA \citep{Adriani2011} and the AMS-02 \citep{AMS02_PROTON, AMS02_HELIUM} experiments, followed by a subsequent softening above around $10$~TeV/n. Various explanations for the spectral hardening have been proposed which are based on physical mechanisms that can affect the CR source spectrum  \citep{Biermann2010, Ohira2011, Yuan2011, Ptuskin2013} and the CR propagation properties in the Galaxy \citep{Blasi2012, Tomassetti2012, Thoudam2014}, as well as explanations based on the presence of nearby sources  \citep{Erlykin2012, Thoudam2012b, Thoudam2013, Bernard2013}. However, a convincing explanation for the origin of the spectral bump is still lacking.

Current explanations for the spectral bump are based on a multi-component origin of CRs involving nearby sources. For instance, \cite{Yue2019} proposed the existence of multiple CR components originating from different populations of distant and nearby sources, while \cite{Malkov2021} proposed the presence of nearby Epsilon Indi or Epsilon Eridani stars as the origin of the bump, and \cite{Li2024} explained it on the basis of the presence of nearby SNRs. In this work, we also present an explanation based on the nearby SNRs, but we bring several improvements over the other existing works. We consider a similar set of potential nearby SNRs that have previously been considered in \cite{Thoudam2012b} to explain the spectral hardening observed at a few hundred GeV. We calculate the CR flux from the nearby SNRs in the presence of a background flux from the distant sources, where we use  updated ages and distances of the nearby SNRs, a consistent spectral index for the CR source spectrum between the background and the nearby flux components, finite sizes of the nearby SNRs in contrast to the commonly assumed point-like approximation, and a time-dependent escape for CRs of different energies from the remnants, unlike other works focusing on the spectral bump.

The effect of source discreteness on the observed spectra of CR nuclei has also been investigated, for example, in \cite{Busching2005} and \cite{Strong2009} on a different context. Studies of the effect on the electron spectrum can be found in \cite{Mao1972} and \cite{Cowsik1979}, and later in \cite{Kobayashi2004, Mertsch2011}, and \cite{Thoudam2012a}

The paper is organized as follows. In Section \ref{section:model}, we describe our model, and present the calculations of the CR fluxes from the nearby and the distant sources. We present the main results in Section \ref{section:results}, the discussion of our findings in Section \ref{section:discuss}, and the conclusion in Section \ref{section:conclude}.

\section{The model}
\label{section:model}
It is widely accepted that SNRs are the major  accelerators of CRs in the Galaxy (e.g., \citealp{ Lagage1983, Axford1994}), which is supported  both by theoretical and observational evidences. Theoretically, it has been established that supernova shocks can accelerate suprathermal particles of the interstellar medium (ISM) to very high energies through the diffusive shock acceleration (DSA) process (\citealt{Axford1977, Bell1978, Blandford1978}). In addition, observational evidence comes from the detection of non-thermal radio and X-ray emissions \citep{Vink2003, Parizot2006} as well as GeV-TeV gamma rays from a number of supernova remnants \citep{Ackermann2013, Abdalla2018}. Based on these evidences, we consider SNRs as the main sources of CRs, at least up to an energy of a few PeVs. 

In our model, we consider two distinct components of the CRs observed at the Earth: a steady background which dominates the observed flux at most of the energies and a time-dependent local component  contributed by the nearby sources (see e.g. \citealp{Thoudam2013}). The background CR component is assumed to originate from distant SNRs  distributed uniformly throughout the Galactic disk. In contrast, the local component is considered to be produced from the nearby SNRs that are within $\sim 1$ kpc of the Earth. They are listed in Table \ref{tab:snrs} along with their estimated ages and distances from the Earth. \citet{Thoudam2006} showed that sources located mainly within $\sim\,1$~ kpc can produce a noticeable variation in the CR flux at the  Earth.

\begin{figure}
%\vspace{24pt}
\includegraphics[width=85mm]{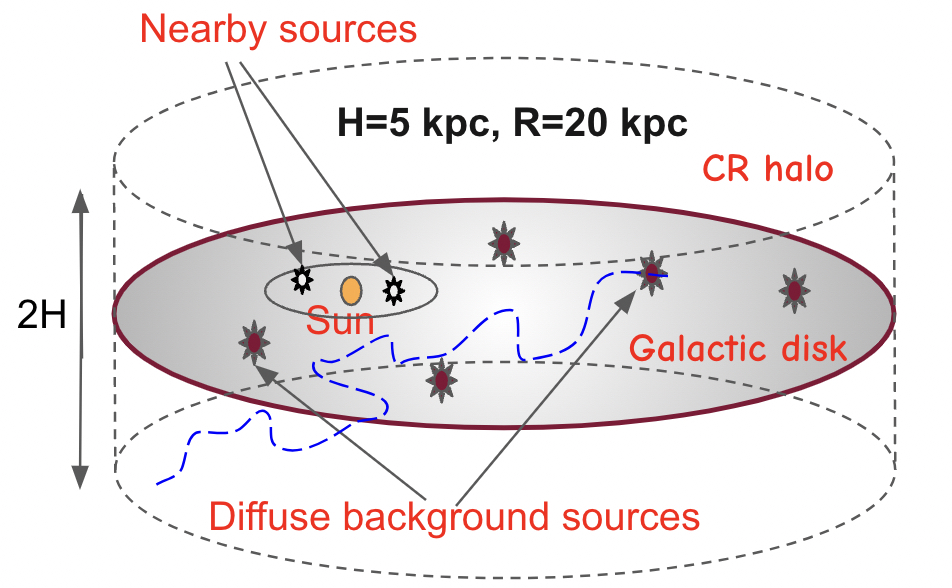}
\caption{Schematic of the distribution of background and nearby SNRs on the Galactic plane. The encircled region around the sun depicts the $1$~kpc size region within which the nearby SNRs considered in this work are located. The blue dashed line illustrates the propagation of CRs from an SNR.} 
\label{fig:galaxy}
\end{figure}

The diffusion region for the background CRs is considered to have a cylindrical geometry with a vertical boundary at $\pm H$, and no boundary in the radial direction (see Figure \ref{fig:galaxy} for the schematic). This is a reasonable assumption as far as the CRs at the Galacto-centric position of the Sun is concerned because majority of them are produced from sources located within a radial distance close to $H$ from the Sun \citep{Thoudam2008}. The precise value of $H$ remains uncertain. Estimates derived from various CR propagation models span a wide range of $\sim 2-12$~kpc \citep{Strong1998, Webber1998}. We consider a typical value of $H=5$ kpc for the present analysis. In contrast, for the local CR component, we assume a diffusion region that is not constrained by any spatial boundary. This assumption is justified because the CR flux from the nearby sources remains largely unaffected by the presence of both the vertical and radial boundaries because of the very short propagation time of CRs to the Earth. The diffusive propagation time for CRs in the Galaxy follows $t_\mathrm{d}\propto r^2/D$, where $r$ is the average distance covered and $D$ is the CR diffusion coefficient. Using this, one can easily check that the CR propagation time to the Earth from a source located 1~kpc away is a factor $0.04$ smaller than the escape time from the Galaxy (considering a halo boundary of $H=5$~kpc).
%For a source located at 1~kpc  (since the sources we consider here, are distributed within $1$ kpc distance from Earth) relative to the escape times from the Galactic boundaries \citep{Thoudam2008}. The propagation time ($t\approx L^2/4D$) for a TeV proton from 1 kpc distance is $\approx 0.2$ Myr (using a diffusion coefficient $D=1.55 \times 10^{28} (E/3 \, {\rm GeV})^{0.5}$ cm$^2$ s$^{-1}$), which is much shorter than the confinement time of $\approx 10$ Myr of CRs.
%
\begin{table}
	\centering
	\begin{tabular}{lccr} % four columns, alignment for each
		\hline
		SNR name & Distance (kpc) & Age (yr)  & Ref.\\
		\hline
		Geminga &  $0.25$ & $3.4 \times 10^5$ & a,b\\
		Loop1 &  $0.17$& $2\times 10^5$ & c\\
		Vela &  $0.30$ & $10^4$ & d \\
            Monogem & $0.30$ & $6.8 \times 10^4$ & e\\
            Cygnus Loop &  $0.73$& $2\times 10^4$ & f\\
            G 114.3+0.3 &  $0.70$& $8\times 10^3$ & g \\
            Vela Junior & $0.70$ & $3.7 \times 10^3$ & h,i\\
            S147 &  $0.90$& $3 \times 10^4$ & j\\
            HB9 & $0.80$& $8 \times 10^3$ & k\\
            HB21 & $0.80$& $1.5\times 10^4$ & l\\
            SN185 & $0.95$& $1.8\times 10^3$ & m\\
            
		\hline
	\end{tabular}
        \caption{List of supernova remnants within a distance of $1$ kpc from the Earth considered in this work. The references are: (a) \citet{Faherty2007}, (b) \citet{Aharonian2023}, (c) \citet{Dickinson2018}, (d) \citet{Sushch2011}, (e) \citet{Cappiello2023}, (f) \citet{Fesen2018}, (g) \citet{Yar-Uyaniker2004}, (h) \citet{Maxted2018}, (i) \citet{Berezhko2009}, (j) \citet{Gvaramadze2006}, (k) \citet{Leahy2007}, (l) \citet{Lazendic2006}, (m) \citet{Hess2018}.}
        \label{tab:snrs}
\end{table}
\subsection{Cosmic rays from the nearby sources}
\label{sec:nearby_sources}
After acceleration at the supernova shocks, CRs eventually escape and undergo diffusive propagation through the ISM. The transport of CRs originating from a single nearby SNR can be described by a time-dependent diffusion equation as,
\begin{eqnarray}
    \nabla \cdot (D\nabla N_\mathrm{p})+Q=\frac{\partial N_\mathrm{p}}{\partial t}\,\, ,
    \label{eq:diffusion_nearby}
\end{eqnarray}
where $N_\mathrm{p}(r, E,t)$ denotes the number density of CR primary particles of kinetic energy per nucleon $E$ at a distance $r$ from the SNR at a given time $t$ after the supernova explosion, $D(E)$ is the diffusion coefficient of CRs in the Galaxy, and $Q(r, E,t)=q(r)q(E)q(t)$ is the source term denoting the CR injection rate per unit volume from the SNR. In Equation \ref{eq:diffusion_nearby}, we neglect CR losses due to the nuclear interaction with the interstellar matter since the time for CRs to reach the Earth from the nearby SNR is expected to be much less than the nuclear interaction time scale. For instance, for a typical interstellar matter density of $n_\mathrm{H}\approx 1$~H cm$^{-3}$, the interaction time for CR protons is $t_\mathrm{pp}\approx(n_\mathrm{H} \sigma c)^{-1} \sim 30$ Myr, where $c$ is the velocity of light. This is larger than the diffusive propagation time $t_\mathrm{d}\approx r^2/4D\sim\,0.3$~Myr for 1~TeV protons to the Earth from a source located at $1$ kpc (see Section \ref{sec:background_sources} for the references to the interaction cross-section $\sigma$ and Section \ref{sec:BC_ratio} for the CR diffusion coefficient $D(E)$). As we focus mainly at high energies in this work, we also neglect effects that are important mostly for CRs at low energies (below $\sim\,10$~GeV) such as re-acceleration, ionization losses and convection by the Galactic wind. In addition, we neglect the production of CR secondaries (if any) inside the remnant from the interaction of the primary CRs before they are released into the ISM. 

In the standard theory of diffusive shock acceleration (DSA) applied to SNRs, charged particles undergo acceleration each time they traverse the supernova shock front \citep{Axford1977, Bell1978, Blandford1978}. During the acceleration, majority of the particles are carried downstream of the shock and are unable to undergo further acceleration, whereas a small fraction diffuses upstream which are recaptured by the expanding shock and continues acceleration. This is true mostly during the free expansion phase of the SNR evolution where the shock moves at a uniform speed with its radius increasing linearly proportional to the time, whereas the displacement of the particles undergoing random walk grows proportional to the square root of the time. However, during the Sedov phase, when the shock slows down, the upstream diffusing particles can start to escape the remnant. Generally, particles are considered to escape the remnant when the upstream diffusion length $D_\mathrm{u}/\upsilon_s$ is larger than the escape boundary $\xi R_\mathrm{s}$, where $D_\mathrm{u}$ is the upstream diffusion co-efficient, $(\upsilon_s, R_\mathrm{s})$ are the velocity and radius of the shock, both function of the SNR age, and the constant $\xi\ll1$ (see e.g., \citealp{Malkov2001, Drury2012}). In the Bohm diffusion limit, where the lowest value of the diffusion coefficient is achieved, $D_\mathrm{u}\propto E/B_\mathrm{u}$, and using  the escape condition, we get $E_\mathrm{esc}\propto B_\mathrm{u}R_\mathrm{s}\upsilon_\mathrm{s}$ for the energy of the escaping cosmic rays. During the Sedov phase, $R_\mathrm{s}\propto t^{0.4}$ and $\upsilon_\mathrm{s}\propto t^{-0.6}$, then the escape energy decreases (weakly) with the time as $E_\mathrm{esc}\propto t^{-0.2}$. If magnetic field amplification is taken into account as suggested \citep{Volk2005, Caprioli2009}, the escape energy can have a stronger time dependence. The magnetic field in the presence of such an amplification scales with the shock velocity as $B_\mathrm{u}\propto \upsilon_\mathrm{s}^d$, where $d$ is a positive constant which can reach a value of 1.5 \citep{Bell2004}. With this, we get the escape energy decreasing with the time as, $E_\mathrm{esc}\propto t^{-(0.2+0.6d)}$. Considering that the magnetic field amplification and the nature of the magnetic turbulence are poorly understood, a simple but reasonable approach is to parameterize the escape energy as  \citep{Gabici2009, Ohira2011, Thoudam2012a},
\begin{equation}
E_{\rm esc} = E_{\rm max}\, \left(\frac{t}{t_{\rm sed}}\right)^{-\alpha},
\label{eq:escape_energy}
\end{equation}
where $\alpha$ is a positive constant. Equation \ref{eq:escape_energy} assumes that the particles with the highest energy $E_{\rm max}$ escape at the onset of the Sedov phase $t_{\rm sed}$, followed by the lower energy particles at later times when the shock becomes weak.

From Equation \ref{eq:escape_energy}, we can write the cosmic-ray escape time as a function of energy as,
%The diffusion upstream is considered to be mostly regulated by the magnetic turbulence generated by the cosmic rays themselves. In the region far ahead of the shock, particle density becomes significantly low and not enough waves can be excited to effectively scatter the particles. Particles reaching this region will no longer be captured by the expanding shock, and they will eventually escape the remnant. Even in the presence of an efficient upstream scattering, particles can escape during the Sedov phase of the SNR evolution as the shock slows down with the SNR age. This sets an escape boundary of the particles diffusing upstream.
\begin{eqnarray}
    t_{\rm esc} = t_{\rm sed}\, \left(\frac{AE}{Ze\rho_m}\right)^{-1/\alpha}\,\,,
    \label{eq:escape_time}
\end{eqnarray}
where $E$ is the kinetic energy per nucleon, $\rho_m=1$~PV is the maximum rigidity \citep{Berezhko1996}, and $\alpha=2.4$ as determined by the spectral fit performed in \cite{Thoudam2013}. It may be noted that the magnetic field amplification level $(d=1.5)$ as suggested by \cite{Bell2004} corresponds to a value of $\alpha=1.1$ for a Bohm-like diffusion. In our model, the  time-dependent escape of CRs is crucial to explain the spectral hardening of protons and heavier nuclei observed at a few hundred GeVs by several experiments. Equation \ref{eq:escape_time} implies that for the same energy/nucleon, heavy nuclei escape at a relatively early time compared to protons, with a time scale that is shorter by a factor of $(A/Z)^{-1/\alpha}$.

As discussed above,  SNRs can effectively confine high-energy particles during the early stage of their evolution. At a later stage, when the shock slows down and cannot efficiently accelerate particles, it can no longer effectively confine the particles. We assume that all the low-energy particles escape the remnant at $10^5$ yr when the shock becomes too weak. In our model, this happens at energies below $\sim 3$~GeV for protons. For the details of the mechanism of particle escape from SNR, see for example \citet{Caprioli2009} and references therein. The CR escape time in our model is then considered as, $T_{\rm esc} = {\rm min}\,\,[t_{\rm esc}(E),10^5\,\, {\rm yr}]$. The corresponding escape radii can be calculated using the age–radius Sedov relation for SNRs as given below \citep{Thoudam2013}:,
\begin{eqnarray}
    R_{\rm esc}=2.5\upsilon_0\,t_{\rm sed}\,\left[\left(\frac{T_{\rm esc}}{t_{\rm sed}}\right)^{0.4}\,-0.6\right]\,\,,
\end{eqnarray}
where $\upsilon_0$ is the initial shock velocity. We consider the SNR to be spherically symmetric. Assuming that CRs are uniformly distributed on the surface of the SNR before they are released, the source term in Equation \ref{eq:diffusion_nearby} is written as,
\begin{eqnarray}
    Q(r,E,t)=\frac{q(E)}{4\pi R_{\rm esc}^2}\,\delta(r-R_{\rm esc})\, \delta(t-T_{\rm esc})\,\,.
\end{eqnarray}
where $q(E)$ is the source spectrum which can be expressed in terms of the total kinetic energy of the particle, $U=AE$, as follows,
\begin{eqnarray}
    q(E)&=&Aq(U)\nonumber\\
    &=& A k \times (U^2+2Um)^{-(\gamma+1)/2}\,(U+m).
    \label{eq:source_spectrum}
\end{eqnarray}
with $m$ representing the rest mass energy of the particle, $\gamma$ the source spectral index, and $k$ a constant related to the CR injection efficiency which is  defined as the fraction $f$ of the supernova kinetic energy of $10^{51}$~ergs injected into a given CR species. The solution of Equation \ref{eq:diffusion_nearby} follows \citep{Thoudam2012a},
\begin{eqnarray}
    N_p(r,E,t)&=&\frac{q(E)R_{\rm esc}}{rA_{\rm esc}\sqrt{\pi D(t-T_{\rm esc})}}\, {\rm exp}\,\left[-\frac{R_{\rm esc}^2+r^2}{4D(t-T_{\rm esc})}\right]\,\nonumber\\
    && \times{\rm sinh} \left[\frac{rR_{\rm esc}}{2D(t-T_{\rm esc})}\right],
    \label{eq:nearby_solution}
\end{eqnarray}
where $A_{\rm esc}=4\pi R^2_{\rm esc}$ is the surface area of the SNR at the instant when CRs of kinetic energy/nucleon $E$ escape the remnant. Equation \ref{eq:nearby_solution} is valid only for time $t\geq T_{\rm esc}$. The CR flux from the source at $t<T_{\rm esc}$ is taken to be zero.

The onset of the Sedov time depends on the initial shock velocity of the SNR ($\upsilon_0$), the initial ejecta mass ($M_{\rm ej}$), and the ambient ISM density ($\rho$), as $t_{\rm sed}\approx (3 M_{\rm ej}/4 \pi \rho)^{1/3} \upsilon_0^{-1}$. Typical values of Sedov time lie in the range $\sim 10^2-10^3$ yr. We consider a uniform value of $t_{\rm sed} = 500$ yr for all of the nearby SNRs \citep{Thoudam2013}. This gives the CR escape time in our model in the range of $T_{\rm esc}=(500-10^5)$~yr depending on the energy, and the corresponding escape radius as   $R_{\rm esc}\sim (5-100)$~pc for a typical shock velocity of $v_0=10^4$~km~s$^{-1}$. We assume all the nearby SNRs to share the same set of model parameters $\lbrace q(E), T_\mathrm{esc}(E), R_\mathrm{esc}(E)\rbrace$. Additionally, in our model, we consider the same CR source index for the nearby and the background sources.

\subsection{Cosmic rays from the background sources}
\label{sec:background_sources}
The flux of the background primary CR component can be calculated from the time-independent (steady-state) propagation equation \citep{Thoudam2013},
\begin{eqnarray}
    \nabla \cdot [D\nabla N_\mathrm{p}(E)]-\eta \upsilon_\mathrm{p}\sigma_\mathrm{p} \delta(z)N_\mathrm{p}(E)=-Q_\mathrm{p}(E),
    \label{eq:diffusion_background}
\end{eqnarray}
where $N_\mathrm{p}(\textbf{\textit{r}},E)$ represents the CR number density with  kinetic energy/nucleon $E$ in cylindrical coordinate $\textbf{\textit{r}}=(r, z)$ with the center at the Galactic center position. The first term of Equation \ref{eq:diffusion_background} represents diffusion, and the second term represents catastrophic loss due to the collisions of CRs with the ISM particles, where $\eta$ is the averaged surface density of interstellar matter in the Galactic disk, $\upsilon_\mathrm{p}$ is the velocity and $\sigma_\mathrm{p}(E)$ is the inelastic collision cross-section of the primary CR particle. We consider a uniform distribution of the background sources, represented by $Q_\mathrm{p}(\textbf{\textit{r}},E)=Sq(E)\delta(z)$, where $S$ is the supernova explosion rate per unit surface area in the Galactic disk and $q(E)$ is the source spectrum defined by Equation \ref{eq:source_spectrum}. We use the same source index for the background and the local components, but allow them to have different CR injection efficiencies considering the uncertainties in the supernova rate in the Galaxy. The solution of Equation \ref{eq:diffusion_background} at $\textbf{\textit{r}}=0$ is given by \citep{Thoudam2013}, 
\begin{eqnarray}
    N_\mathrm{p}(0,0,E)=\frac{RSq_\mathrm{p}(E)}{2D_\mathrm{p}}\,\int_0^\infty\,\frac{J_1(KR)}{K\, {\rm coth} (KH)+\frac{\eta \upsilon_\mathrm{p} \sigma_\mathrm{p}}{2D_\mathrm{p}}}\, dK,\nonumber\\
    \label{eq:solution_background}
\end{eqnarray}
where, $J_1$ denotes the Bessel function of first order and $R$ is the radial extent of the source distribution which is taken to be $20$ kpc. The collision cross-section has been taken from \cite{Thoudam2013} which is based on the cross-sections reported in \cite{Kelner2006} for the protons and \cite{Letaw1983} for the heavier nuclei.  %IS NOT CORRECT. For protons, we use the cross-section from kelner et al., and for heavier nuclei from Letaw et al. We use Heinbach1995 for the primary to secondary production.]}.

Primary CRs produce secondary particles due to their interaction (spallation) with the ISM particles during their propagation in the Galaxy. The secondary CR production rate can be calculated as,
\begin{eqnarray}
   Q_s(\textbf{\textit{r}},E)=\int_E^{\infty} \eta \upsilon_\mathrm{p}N_\mathrm{p}(
\textbf{\textit{r}},E^{\prime})\delta(z) \frac{d}{dE^{\prime}}\sigma_\mathrm{ps}(E,E^{\prime})dE^{\prime},\nonumber\\
\label{eq:source_term_secondary}
\end{eqnarray}
where the suffix `s' indicates secondary species, $N_\mathrm{p}p$ is the primary CR number density, and $d \sigma_\mathrm{ps}(E, E^{\prime})/dE^{\prime}=\sigma_\mathrm{ps}\delta(E^{\prime}-E)$ denotes the differential production cross-section of a secondary nucleus with energy per nucleon $E$ from a primary nucleus of energy per nucleon $E^{\prime}$ and $\sigma_\mathrm{ps}$ is the total production cross-section of the secondary. Equation \ref{eq:source_term_secondary} then becomes
\begin{eqnarray}
    Q_\mathrm{s}(\textbf{\textit{r}},E)=\eta \upsilon_\mathrm{p} \sigma_\mathrm{ps} N_\mathrm{p}(
\textbf{\textit{r}},E)\delta(z).
\label{eq:source_s}
\end{eqnarray}

The density of the secondary CRs can be obtained using a similar transport equation that describes the primaries (Equation \ref{eq:diffusion_background}), but with the source term replaced by Equation \ref{eq:source_s}. Their density at $\textbf{\textit{r}}=0$ is given by \citep{Thoudam2013},
\begin{eqnarray}
    N_\mathrm{s}(0,0,E)=\eta \upsilon_\mathrm{p} \sigma_\mathrm{ps} N_\mathrm{p}(0,0,E)\frac{R}{2D_\mathrm{s}} \times \nonumber \\
    \int_0^{\infty}\frac{J_1(KR)\,dK}{K\, {\rm coth} (KH)+\frac{\eta \upsilon_\mathrm{s} \sigma_\mathrm{s}}{2D_\mathrm{s}}},
    \label{eq:s_to_p}
\end{eqnarray}
where $N_\mathrm{s}(0,0,E)$ is given by Equation \ref{eq:solution_background}. For the calculation of the secondary particles, it should be mentioned that we only consider primaries from the CR background, but not the primaries originating from the nearby SNRs as they do not get enough time to interact  with the ISM before reaching the Earth as discussed in Section \ref{sec:nearby_sources}.

Our study focuses only on boron nuclei as the secondary CR species. They are known to produce mainly from the spallation of the $^{12}$C and $^{16}$O primaries in the Galaxy. These primaries, upon interaction, generate ($^{11}$B,$^{10}$B) and ($^{11}$C,$^{10}$C) isotopes. The latter subsequently decays into ($^{11}$B,$^{10}$B). For our calculations, we take the secondary production cross-sections from \citep{Heinbach1995}.

\subsection{Secondary-to-primary ratio from the background cosmic rays}
\label{sec:BC_ratio}
\begin{figure}
%\vspace{24pt}
\includegraphics[width=85mm
]{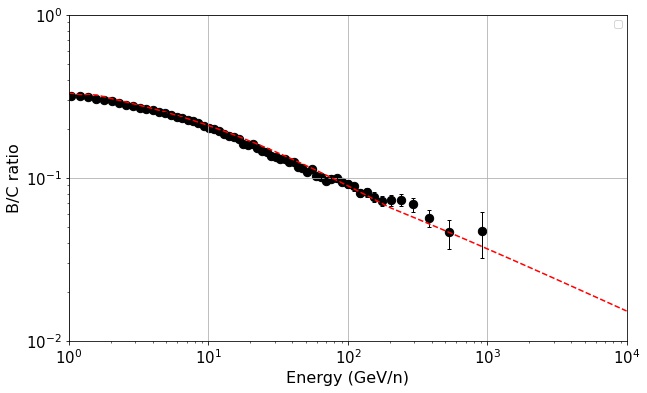}
\caption{Boron-to-carbon ratio for the background cosmic rays (dashed line). Data points are from the AMS-02 experiment \citep{AMS02_BC_RATIO}.}% and excluding the nearby sources. The cyan dashed line shows our model prediction, where we include both the nearby and background contributions for the primaries but only the background contribution for secondaries (see Eq. \ref{eq:BbyC_inclnearby}).} 
\label{fig:BC_ratio_2}
\end{figure}

The secondary-to-primary ratio for the background CRs can be calculated from Equation \ref{eq:s_to_p}. The ratio, $N_\mathrm{s}/N_\mathrm{p}\propto 1/D_\mathrm{s}$, is used to determine the diffusion coefficient in the Galaxy. For the present study, we model $D(E)$ in the form of,
\begin{eqnarray}
    D(E)&=&D_0\left(\frac{\upsilon}{c}\right)\,\left(\frac{E}{E_0}\right)^{\delta},\,\,\,\,\,\, E\leq E_1 \,\nonumber\\
    &=&D_0\left(\frac{\upsilon}{c}\right)\,\left(\frac{E}{E_1}\right)^{\delta_1}\,\left(\frac{E_1}{E_0}\right)^{\delta},\,\, E>E_1 \,\,
    \label{eq:diffusion_index}
\end{eqnarray}
where, again, $E$ is the kinetic energy per nucleon, $\upsilon$ is the particle velocity, $c$ is the velocity of light, $D_0$ is the diffusion constant, and $\delta$ ($\delta_1$) are the diffusion indices below  (above) the break energy $E_1$. Optimizing Equation \ref{eq:diffusion_index} to the boron-to-carbon ratio (B/C) data from the AMS-02 experiment \citep{AMS02_BC_RATIO}, we obtain $D_0=1.55\times 10^{28}$ cm$^2$s$^{-1}$, $\delta=0.54$, $\delta_1=0.40$ and $E_1=200$ GeV/n. $E_0$ is fixed at $3$~GeV/n in the procedure. The result is shown in Figure \ref{fig:BC_ratio_2} (dashed line). In the pure diffusion model of CR propagation in the Galaxy such as the one considered in this work, the slight hardening observed in the B/C data above $\sim\,100$~GeV/n requires a change in the diffusion index towards smaller values at high energies. For simplicity, this is often incorporated as a break in $D(E)$ around $100-300$~GeV/n \citep{Genolini2017}. Similar behavior of $D(E)$ has also been introduced to explain the hardening in the cosmic-ray primary spectra above $\sim\,200$~GeV/n on the assumption that there exists a respective break in the turbulence spectrum in the Galaxy \citep{Blasi2012}. However, it has been recently argued that a break in the turbulence spectrum is unlikely to introduce a corresponding significant break in the particle spectrum because the ``exact" cyclotron wave-particle resonance condition, $p=eB/ck\cos\theta$, also involves the particle's pitch angle $\theta$, in addition to the wave number $k$ and the particle momentum $p$ \citep{Malkov2024}. Here, $B$ is the ISM magnetic field, $e$ the charge of an electron, and $c$ the velocity of light. The exact resonance condition allows particles with the same momentum but with different pitch angles to interact with waves of different wave numbers. This will eventually smear out the effect of the break in the turbulence spectrum on the particle spectrum. Moreover, even if the spectral break in the particle spectrum is formed, the break feature can be smoothened by the diffusion in momentum space which particles experience during their propagation through the Galaxy, unless the break is formed not so far away from Earth \citep{Malkov2024}. For the work presented here, which focuses on explaining the spectral bump (along with the spectral hardening above $\sim\,200$~GeV/n) based on local sources, the break in $D(E)$ introduced in Equation \ref{eq:diffusion_index} will not produce significant effect on the main results. It is also worth mentioning that models involving CR re-acceleration in the Galaxy do not require such a break in $D(E)$ in order to explain the B/C data (see e.g., \citealt{Ptuskin1994, Thoudam2014}). %This set of parameters for $D(E)$ of the three different optimization models considered in this work is described in detail in Section \ref{sec:Diff_scenario}. Note that these values, in particular $D_0$, which are obtained purely based on the background CRs for now, will be further %optimized modified by adding the contribution of CRs from the nearby sources (see Section \ref{sec:Diff_scenario}). 

It may be noted that recent measurements of B/C by the CALET \citep{Adriani2022_BC} and DAMPE \citep{Dampe_BC} experiments show an indication of flattening of the ratio above $\sim 1$ TeV. Such a flattening may be an effect of the re-acceleration of CRs by strong shocks associated with young SNRs during their propagation in the Galaxy \citep{Wandel1987, Blasi2017} or due to the production of secondaries inside the source region from the interaction of primary nuclei with the local matter \citep{Berezhko2003, Cowsik2010}. Neither of these mechanisms will affect the background CR primaries. However, the shape of the background secondaries can be affected, particularly at the high energies, due to the generation of an additional component of secondaries with a source spectrum similar to that of the primaries which is  much flatter than that of the secondaries produced in the ISM. In the present study, we neglect the flattening in B/C at TeV energies since it does not affect the major results presented here, which focus on the spectra of the primary CRs and the all-particle spectrum.

\subsection{Additional model parameters}
\label{sec:parameters}
Considering that the interstellar matter is distributed mostly within the thin Galactic disk, we use the surface matter density instead of the actual number density for our calculations. The average surface density has been taken as  $\eta=5.17\times10^{20}$ H atoms cm$^{-2}$ \citep{Thoudam2013}, estimated using the atomic and molecular hydrogen observations. The value represents the average over a circle of $5$~kpc radius from the position of the Sun. We add $10\%$ to this value to take into account the contribution of helium atoms in the ISM. The supernova surface density in Equation \ref{eq:solution_background} is taken as $S=7.7$ Myr$^{-1}$ kpc$^{-2}$,  which corresponds to a supernova rate of $0.98$ per century in the Galaxy. This value of the supernova rate is consistent with the $1.9\pm1.1$ per century, as inferred by \cite{Diehl2006}. Additionally, the effect of solar modulation is included by considering the force-field approximation with a modulation parameter $\phi$ \citep{Gleeson1968}. We take $\phi=400$~MV as this value is found to produce an overall good fit to the data below $\sim\,10$~GeV/n for the different CR species considered in this work, as demonstrated in Section \ref{element_spectra}. %This value has been determined by fitting our calculated proton and helium spectra with the lower energy data points measured by AMS02 \citep{Aguilar2018}. % and PAMELA \citep{Adriani2011}.

\begin{figure}
%\vspace{24pt}
\includegraphics[width=85mm, height=110mm
]{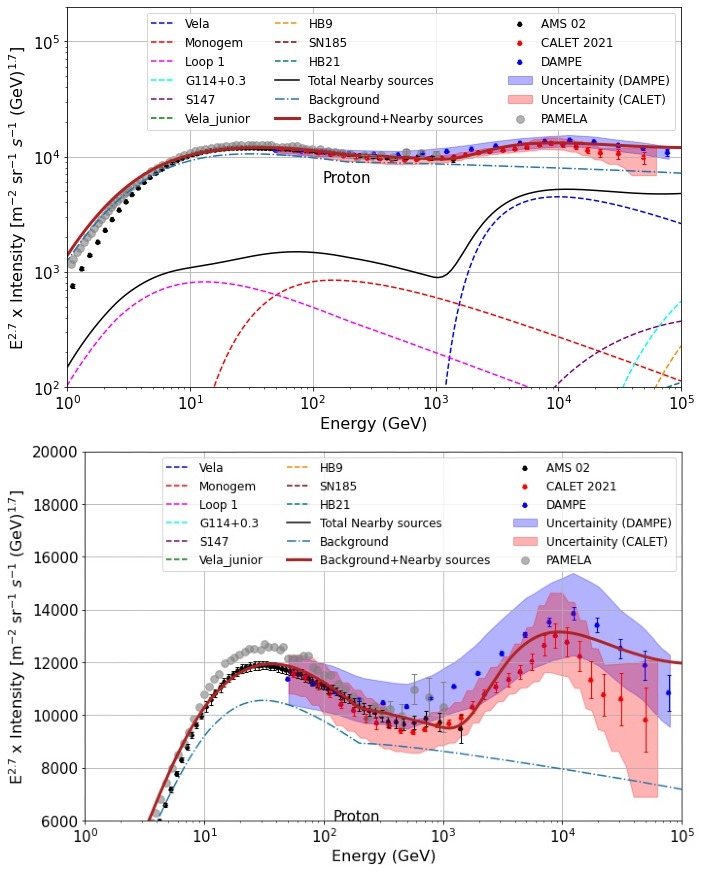}
\caption{Top panel: Proton spectrum in log-log plot. Bottom panel: Same spectrum in the log-linear plot. The dashed lines show the individual contribution from the nearby SNRs listed in Table \ref{tab:snrs}, the black solid line shows their total contribution, and the dash-dotted line represents the background component. The red solid line shows the calculated total (nearby+background) proton flux at the Earth. Data points are taken from AMS-02 \citep{AMS02_PROTON}, PAMELA \citep{Adriani2011}, CALET \citep {Adriani2022p}, DAMPE \citep{An2019} experiments. The individual contributions of the nearby SNRs are not visible in the bottom panel.}
\label{fig:proton}
\end{figure}

\section{Results on the cosmic-ray spectra}
\label{section:results}
In this section, we present the spectra of the individual CR elements obtained with our model, and demonstrate that the presence of the nearby SNRs can be responsible for the observed spectral bump between $\sim\,1$~TeV and 100~TeV in the proton and the helium spectra. We also obtain the all-particle CR spectrum from our model, and compare it with the observed data up to a few times $10^7$~GeV.
\subsection{Individual element spectra}
\label{element_spectra}

\begin{figure}
%\vspace{24pt}
\includegraphics[width=85mm, height=110mm
]{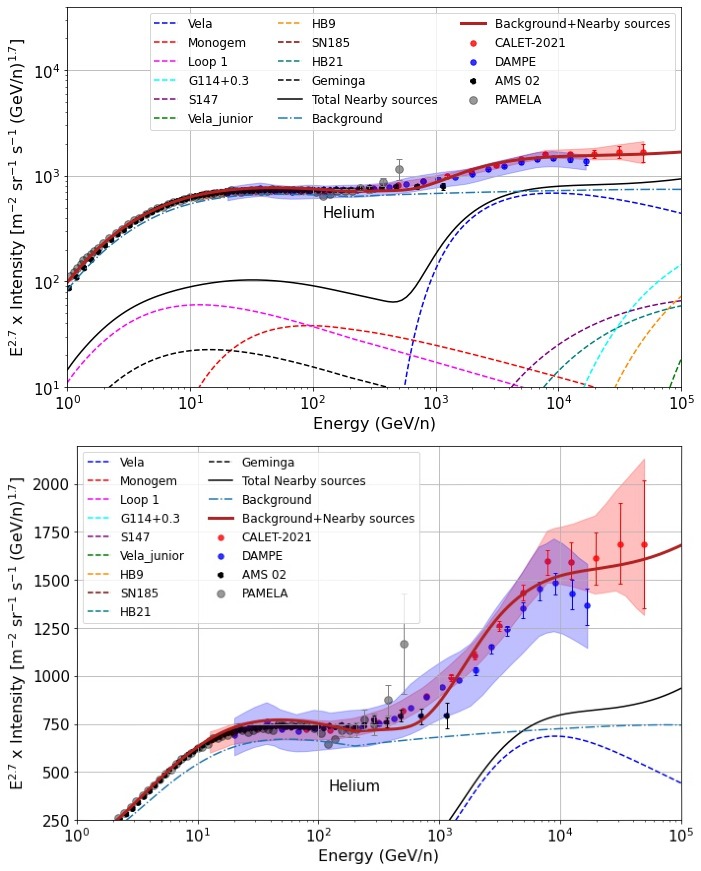}
\caption{Top panel: Helium spectrum in log-log plot. Bottom panel: Same spectrum in the log-linear plot. The dashed lines show the individual contribution from the nearby SNRs listed in Table \ref{tab:snrs}, the black solid line shows their total contribution, and the dash-dotted line represents the background component. The red solid line shows the calculated total (nearby+background) helium flux at the Earth. Data points are taken from AMS-02 \citep{AMS02_HELIUM}, PAMELA \citep{Adriani2011}, CALET \citep{Adriani2023}, DAMPE \citep{Alemanno2021} experiments. Among the nearby SNRs, only the contribution of Vela is visible in the bottom panel.}
\label{fig:helium}
\end{figure}

\textit{Proton and helium spectra:} Figures \ref{fig:proton} and \ref{fig:helium} show the predicted spectra at the Earth for the protons and helium nuclei respectively. The upper panels of the figures present the spectra in the log-log scale which shows the background contribution (dash-dotted line) as well as the individual contribution of the nearby SNRs. The black solid line represents the total contribution from the nearby SNRs, and the red solid line represents the total background plus nearby contribution. The lower panels of the figures show the same spectra as in the upper panels, but in the log-linear scale. In our calculation, the source index $\gamma$ and the CR injection efficiency $f$ are kept as model parameters. We allow the individual CR elements to have different values of $(\gamma, f)$ which are optimized based on the observed individual elemental spectra. However, for a given CR species, we constrain the background and the nearby sources to have the same value of $\gamma$. In addition, we consider all the nearby sources to have the same values of $\gamma$ and $f$ for a given CR species. For the protons (Figure \ref{fig:proton}), we obtain the best-fit values as  $\gamma=2.34$, $f=16\%$ for the background component, and $f=25\%$ for the nearby SNRs. For the helium nuclei (Figure \ref{fig:helium}), we find $\gamma=2.28$, $f=1.52\%$ for the background, and $f=3.1\%$ for the nearby component. We also find that choosing a slightly different value of $D(E)$ for the nearby sources improves the fit quality, especially in the spectral bump region at $\sim\,1-100$~TeV/n observed by the CALET and the DAMPE experiments. For this reason, we choose $D_0= 10^{28}$ cm$^2$s$^{-1}$, $E_0=3$ GeV, $\delta=\delta_1=0.54$ (i.e without break at $200$ GeV) for the CR propagation from the nearby SNRs. For the background component, we use the value of $D(E)$ as described in Section \ref{sec:BC_ratio}. This difference in the diffusion properties is possible if the local region has magnetic turbulence different from the Galactic average experienced by the background CRs in the Galaxy.
%{\color{red} The propagation scenario for the background CRs and nearby CRs can be different. Keeping this in mind we use two different diffusion index for background and nearby in order to explain the spectral bump. In this work, we do not focus on data below $10$ GeV. In the low-energy region, various physical phenomena such as solar modulation, ionization losses, and the effects of convection due to Galactic winds play significant roles, which are not accounted for in our simplified model. Therefore, our analysis is restricted to energies above $10$ GeV.}
%

In our model, the observed spectral bump in the proton and helium spectra at $\sim\,1-100$~TeV/n is explained as a result of the contribution from the nearby SNRs, in particular the Vela supernova remnant. Below $\sim\,1$~TeV/n, the nearby component contributes significantly less compared to the background flux. The steep rise in the CR flux from Vela at $\sim\,1$~TeV/n is mainly due to the energy-dependent CR escape mechanism implemented in our model. CRs below $\sim\,1$~TeV/n are mostly still confined with the remnant. There is also an additional effect due to the slow diffusion of CRs at low energies making them unable to reach the Earth within the given time. On the other hand, CRs above $\sim\,10$~TeV/n, being released at an earlier stage and undergoing faster diffusion, have already passed by the Earth and generates a steep power-law spectrum. It is known that high-energy particles whose diffusion radius, $r_\mathrm{d}\sim\sqrt{D(t-T_\mathrm{esc})}$, is much larger than the distance to the observer, produce a  spectrum that follows $E^{-(\gamma+3\delta_\mathrm{1}/2)}$ \citep{Thoudam2012b}. SNRs such as Loop1 and Monogem dominate the nearby contribution below $\sim\,1$~TeV/n, while others such as Vela Junior, SN185, and HB9 show their contribution only above $\sim\,10$~TeV/n, but they remain subdominant with respect to the contribution from Vela.

The nearby SNRs contribution looks similar between the proton and the helium spectra, except that the helium results are slightly shifted towards lower energies. This difference is primarily due to the early escape time of helium nuclei relative to the protons at the same energy per nucleon. In fact, in our model, all the nuclei heavier than protons are released earlier by a factor of $(A/Z)^{-1/\alpha}$ with respect to the protons (see Equation \ref{eq:escape_time}).

During the time cosmic rays are confined inside the remnant, particles can gain energy due to the second-order Fermi acceleration process resulting from their interaction with the magnetic turbulence. The energy gain depends on the level of the magnetic scattering and the confinement time of the particles. Low-energy particles suffer more scattering, and they escape  at a later stage of the SNR evolution compared to the high-energy particles. Therefore, the energy gain due to the second-order Fermi process is expected to affect mostly particles at low energies. In the following, we present an estimate of the energy gain for protons and its effect on the spectrum at the Earth, considering the Vela supernova remnant as an example.

Cosmic rays remain confined inside the remnant as long as their upstream diffusion length is less than the escape length, as discussed in Section \ref{sec:nearby_sources}. Under this condition, the upstream spatial diffusion coefficient at the time particles of energy $E$ escape the remnant is determined as $D_\mathrm{u}(E)=\xi R_\mathrm{s}(t)\upsilon_\mathrm{s}(t)$, where $\xi=0.1$ \citep{Drury2011}, and $t=t_\mathrm{esc}(E)\propto E^{-1/\alpha}$ as given by Equation \ref{eq:escape_time}. For the value of $\alpha=2.4$ adopted in this work, $R_\mathrm{s}\propto E^{-0.17}$ and $\upsilon_\mathrm{s}\propto E^{0.25}$. This gives $D_\mathrm{u}(E)\propto E^{0.08}$, which shows a very weak dependence on the escape energy. In other words, the upstream diffusion coefficient at an early time when a high-energy particle escapes is more or less similar to the value when a low-energy particle escapes at a later time. This can be understood from our choice of $\alpha=2.4$ in the escape time $t_\mathrm{esc}(E)$ which incorporates a strong magnetic field amplification where the magnetic field scales with the shock velocity as $B_\mathrm{u}\propto\upsilon_\mathrm{s}^d$, with $d$ as a positive constant (see e.g., \citealp{Bell2004}). At the time when high-energy particles of energy $E_\mathrm{1}$ escape, the shock velocity is large, producing a large $B_\mathrm{u}$, and a corresponding small $D_\mathrm{u}(E_\mathrm{1})$. On the other hand, when low-energy particles with energy $E_\mathrm{2}$ start to escape, the shock has already slowed down, lowering the magnetic field amplification, and producing a large $D_\mathrm{u}(E_\mathrm{2})$ which turns out to be comparable to the value of $D_\mathrm{u}(E_\mathrm{1})$. The spatial diffusion coefficient in the downstream, is expected to be smaller as the magnetic field downstream is compressed. We take it as $D_\mathrm{d}=D_\mathrm{u}/4$, where, for simplicity, we assume a compression factor of $4$ which corresponds to strong shocks\footnote{The actual compression factor that is required for our calculation will be slightly less than $4$ as particle escape happens when the shock starts to slow down.}. Then, using the relation between the spatial and the momentum diffusion coefficients, $D_\mathrm{d}D_\mathrm{p}\approx (1/9)p^2V_\mathrm{A}^2$, where $p=E/c$ is the particle momentum and $V_\mathrm{A}$ is the velocity associated with the motion of the magnetic turbulence \citep{Thornbury2014}, the momentum diffusion coefficient $D_\mathrm{p}$ for the cosmic rays in the downstream is calculated. The corresponding gain in momentum during the time cosmic rays are confined downstream is determined from $\Delta p\sim\sqrt{D_\mathrm{p}t_\mathrm{esc}}$, where we assume $D_\mathrm{p}$ constant over time\footnote{In our estimate, we neglect the variation of the spatial diffusion coefficient $D_\mathrm{d}$ (and hence, of $D_\mathrm{p}$) over time. In reality, the value of $D_\mathrm{d}$ (or $D_\mathrm{p}$) is expected to increase (or decrease) with time as the level of the magnetic field amplification goes down with the SNR age.}. This gives an energy gain that decreases with energy as $\Delta E/E\propto E^{-0.25}$, where we have used $\Delta E=\Delta p c$. For $V_\mathrm{A}=100$~km/s, we get $\Delta E/E\sim 8.5\%$ at $1$~TeV, which comes down to $\sim 5.0\%$ at $10$~TeV. When applied to the nearby SNRs, the energy gain will allow particles to escape at a slightly earlier times than they would do in the absence of the energy gain, according to Equation \ref{eq:escape_time}. This, together with the effect of the energy dependent diffusion in the ISM, will shift the CR spectrum at the Earth from a nearby SNR towards lower energies. When applied to the Vela supernova remnant, we have found an increase in its proton flux at the Earth by a factor of $\sim 1.9$ at 1~TeV and $\sim 1.1$ at 10~TeV with respect to the flux in the absence of the second-order Fermi acceleration process shown in Figure \ref{fig:proton}. Although this increase at $\lesssim 1$~TeV looks significant for Vela, the effect on the total proton flux (background plus nearby components) will not be  significant as the nearby SNRs contribute less than $\sim 10\%$ of the total flux below $1$~TeV.

\begin{figure}
%\vspace{24pt}
\includegraphics[width=85mm, height=160mm
]{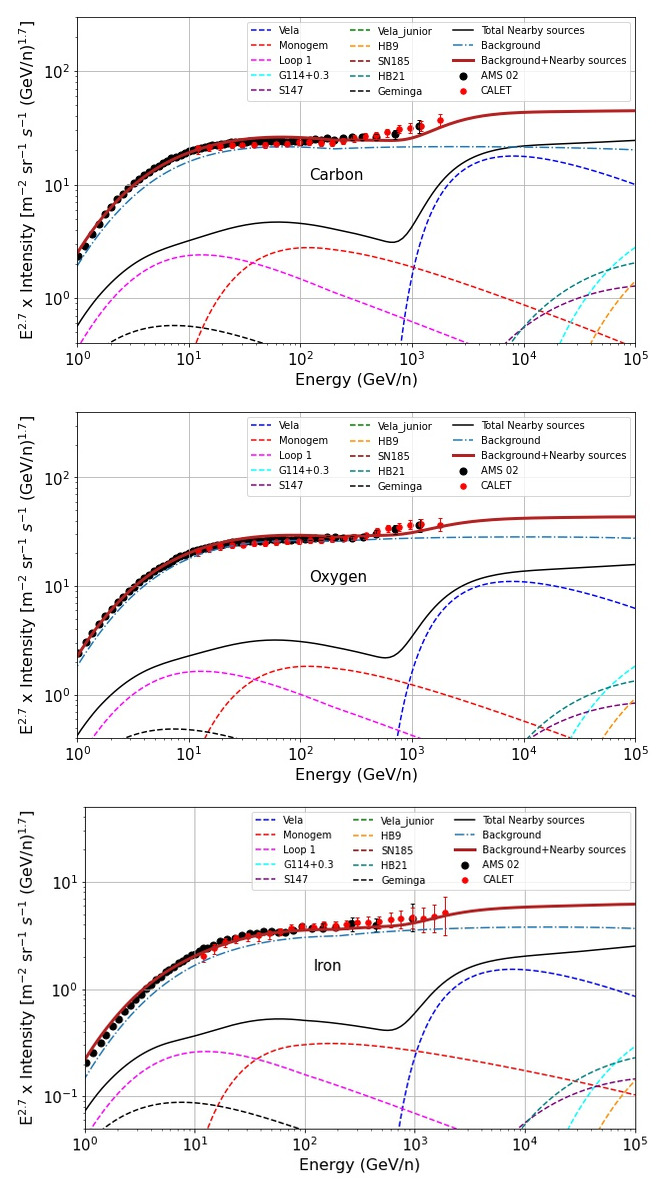}
\caption{Cosmic-ray energy spectra of the carbon (top), oxygen (middle), and iron (bottom) nuclei.  Dash-dotted line: background spectrum, black solid line: total nearby component, dashed lines: individual nearby SNRs, thick-solid maroon line: total nearby plus background. The CALET \citep{Adriani2020, Adriani2021c} data points have been shifted in energy by $+8\%$ for carbon and oxygen, and $+6\%$ for iron to minimize the systematic offset with respect to the AMS-02 data \citep{Aguilar_2017, Aguilar2021}.} 
\label{fig:carbon_oxygen_iron}
\end{figure}

\begin{figure}
%\vspace{24pt}
\includegraphics[width=85mm, height=160mm
]{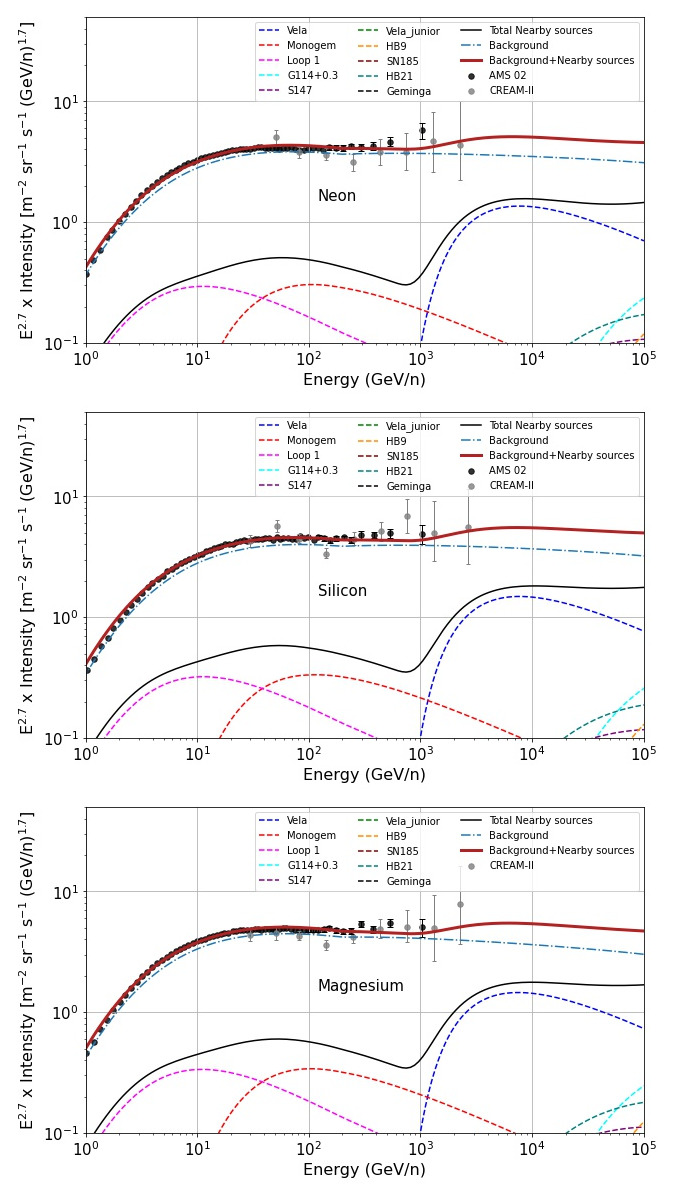}
\caption{Cosmic-ray energy spectra of the neon (top), silicon (middle), and magnesium (bottom) nuclei. Dash-dotted line: background spectrum, black solid line: total nearby component, dashed lines: individual nearby SNRs, thick-solid maroon line: total nearby plus background. The CREAM \citep{Ahn2009b} data points have been shifted in energy by $+11\%$ for neon, silicon, and magnesium to minimize the systematic offset with respect to the AMS-02 data \citep{Aguilar2020}.} 
\label{fig:neon_silicon_magnesium}
\end{figure}

\textit{Heavier elements spectra:} The spectra for the heavier elements are shown in Figure \ref{fig:carbon_oxygen_iron} for carbon, oxygen, and iron nuclei, and in Figure \ref{fig:neon_silicon_magnesium} for neon, magnesium, and silicon nuclei. For these elements, the spectra are shown only in the log-log scale. The effect of the nearby SNRs looks similar to those found in the proton and helium spectra, showing a bump-like feature at $\sim\,(1-100)$~TeV/n. Except for the source index $\gamma$ and the CR injection fraction $f$, all other model parameters remain the same as in the calculation for the proton and helium spectra. Table \ref{tab:fitting_parameters} gives the values of $\gamma$ and $f$ (for the background and the nearby SNRs) for the different elements used in our calculation.

Similarly to the proton and helium results, Vela gives the most significant contribution above $\sim\,1$~TeV/n also in the case of heavier nuclei, producing a signature of spectral hardening at TeV energies. This is in agreement with the observations from the AMS-02 and the CALET experiments. CREAM measurements show large uncertainties at these energies, although a similar trend of spectral hardening seems to be present. Future sensitive measurements at higher energies can provide a crucial check of our prediction for the heavy elements.

\begin{figure}
%\vspace{24pt}
\includegraphics[width=85mm, height=60mm,
]{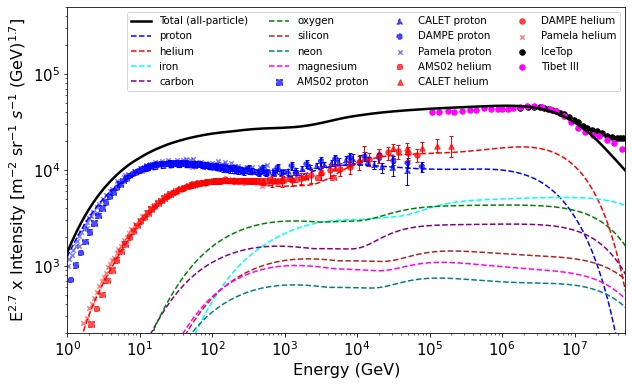}
\caption{All-particle cosmic-ray spectrum predicted by our model (black solid line). Each dashed line represents the spectrum of the individual element which is the sum of the background and the contribution from the nearby SNRs. Individual elemental data are shown only for protons and helium nuclei (same as in Figures \ref{fig:proton} and \ref{fig:helium}). The all-particle data are from IceTop \citep{Aartsen2013} and Tibet III \citep{Amenomori2008} experiments.} 
\label{fig:all_particle}
\end{figure}

\subsection{All-particle spectrum}
We calculate the all-particle CR spectrum by combining the spectra of the different elements shown in Figures~\ref{fig:proton}-\ref{fig:neon_silicon_magnesium}. The result is shown in Figure \ref{fig:all_particle} (black solid line) and compared with the available measurements from IceTop \citep{Aartsen2013} and Tibet III \citep{Amenomori2008} experiments. The dashed lines represent the spectra of the individual elements which is the sum of the contribution from the background and the nearby sources. In order to reproduce the knee feature in the observed all-particle spectrum, we consider an exponential cut-off in the source spectrum at $E_\mathrm{c}=4\times 10^6 Z$ GeV in our calculation, where $Z$ is the charge number of the element. The deficit in our model prediction above $\sim\,10^7$~GeV possibly indicates the presence of a second or additional galactic component \citep{Thoudam2016, Bhadra2024} over the regular sources considered in this work.

\begin{table*}
    \centering
    \begin{tabular}{|c|c|c|c|}
    \hline
    \textbf{Elements} & \textbf{Spectral index} & \multicolumn{2}{|c|}{\textbf{Injection fraction} ($f\times 10^{49}\,\mathrm{ergs}$)}\\
    \cline{3-4}
    & ($\gamma$) & Background sources & Nearby SNRs\\
    \hline \hline
    Proton & 2.34 & 16.0 & 25.0\\
    Helium & 2.28 & 1.52 & 3.10\\
    Carbon & 2.32 & 0.056 & 0.082\\
    Oxygen & 2.31 & 0.062 & 0.070\\
    Iron & 2.32 & 0.010 & 0.010\\
    Silicon & 2.35 & 0.014 & 0.014\\
    Neon & 2.35 & 0.012 & 0.012\\
    Magnesium & 2.38 & 0.014 & 0.016\\
    \hline
    \end{tabular}
    \caption{Values of the source spectral index $\gamma$ and the cosmic-ray injection efficiency $f$ for the different cosmic-ray elements for the background and the nearby SNRs.}
    \label{tab:fitting_parameters}
\end{table*}

\section{Discussion}
\label{section:discuss}
\textit{Explanation of the spectral bump:} In this study, we have shown that the spectral bumps of the cosmic-ray protons and helium nuclei in the TeV region, recently observed by the CALET and the DAMPE experiments, can be explained as an effect of the contribution from nearby SNRs, especially the Vela supernova remnant. In our model, the spectral bump is explained mainly as a result of the  low-energy cut-off due to the energy-dependent escape of CRs from the SNRs and a high-energy fall-off of the spectrum due to the energy-dependent propagation of CRs in the Galaxy. These results are in agreement with the earlier findings presented in \cite{Thoudam2012b, Thoudam2013} in the context of the observed spectral hardening at $\sim\,(200-300)$~GeV/nucleon. In contrast, in a recent analysis, \cite{Li2024} identified other nearby sources such as Geminga, Monogem, and Loop I as the main contributors for the observed spectral bump. Their calculation differs from our approach in that we consider a more realistic energy-dependent escape of particles from the sources while they assume an energy-independent burst-like injection. Moreover, we consider a more consistent approach by fixing the same source spectral index between the nearby SNRs and the background sources, while \cite{Li2024} allows the source index to vary between the  nearby SNRs and also with respect to the background component. Our model also reproduces quite well the observed spectra of the heavier elements up to iron, and at the same time, predicts a spectral bump for the heavier elements similar to that of the protons and helium nuclei which can be tested with future observations. %In addition, our model reproduces the observed all-particle spectrum up to $\sim\,10^8$~GeV when combined with a second Galactic component originating from compact star clusters \citep{Bhadra2024}.

\textit{Comparison with other existing models for the spectral hardening:} The origin of the spectral bump is most likely connected with the spectral hardening observed at $\sim\,(200-300)$~GeV/nucleon. Several models have tried to explain the spectral hardening, but not all the models can easily explain the bump. For instance, explanation based on a break in the diffusion coefficient \citep{Tomassetti2012, Blasi2012} or on a hardened source spectrum \citep{Yuan2011, Ptuskin2013} as well as model based on a global re-acceleration of CRs by weak shocks in the Galaxy \citep{Thoudam2014} successfully reproduce the spectral hardening, but they cannot explain the observed bump unless multiple population of CR sources are invoked (e.g., \citealp{Zatsepin2006, Yue2019}). On the other hand, models based on the presence of nearby sources such as the one presented in this work (see also \citealp{Thoudam2012b, Thoudam2013}) or potential nearby CR re-acceleration sites as proposed in \cite{Malkov2021} can explain the spectral hardening and the bump at the same time. In our model, these features are generated by CRs from the Vela supernova remnant.

\begin{figure}
%\vspace{24pt}
\includegraphics[width=85mm
]{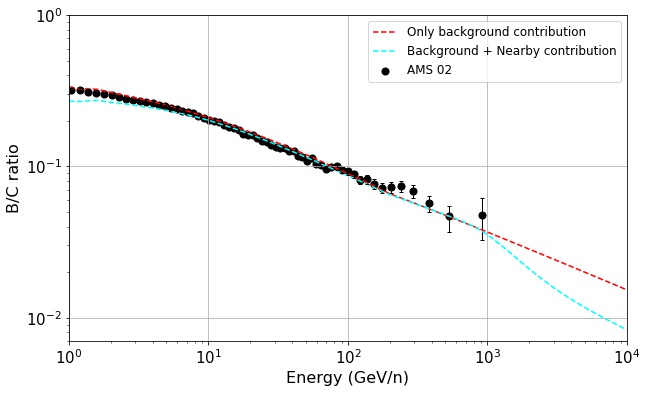}
\caption{Boron-to-carbon ratio in the presence of the nearby SNRs (cyan dashed line). The red dashed line shows our model prediction considering only the background sources (same as shown in Figure \ref{fig:BC_ratio_2}). Data are from the AMS-02 experiment  \citep{AMS02_BC_RATIO}.} 
\label{fig:BC_ratio_3}
\end{figure}

\textit{Effect of nearby sources on the B/C ratio:} The presence of nearby sources directly affects the secondary-to-primary ratios at the energies where their contribution to the primary CR spectra is significant. Secondary spectra remain unaffected as CR primaries from the nearby sources have to travel far distances in the Galaxy by the time they interact with the ISM and produce secondaries. This leads to a negligible contribution of the secondaries produced by the CR primaries from the nearby sources with respect to that produced by the background primaries. The effect will be a suppression in the secondary-to-primary ratios at the energy range where the nearby sources show significant contributions in the primary spectra. This is visible in Figure \ref{fig:BC_ratio_3} (cyan line) at energies above $\sim 1$~TeV/n. However, at these energies, the ratio can also be strongly affected by the re-acceleration of secondary CRs by strong shocks in the Galaxy or by the contribution of additional CR secondaries  from the interaction of CR primaries inside the sources, neither of which are included in the present study, as discussed in Section \ref{sec:BC_ratio}. Therefore, in reality, the steepening in the ratio at high energies caused by the presence of the nearby sources is expected to remain buried under other dominant effects mentioned above, making it hard to observe.

\textit{Effect of nearby sources on cosmic-ray anisotropy:} The contribution of the nearby sources is also expected to produce some level of CR anisotropy at the Earth. For a single source dominating the CR flux at a given energy, the total anisotropy $\delta$ can be  calculated as (e.g., \citealt{Thoudam2007}), 
\begin{equation}
    \delta(E)=\frac{I_{\mathrm{m}}}{I_{\mathrm{T}}}\delta_{\mathrm{m}}
\end{equation}
where $I_{\mathrm{m}}$ denotes the CR intensity from the dominant source at energy $E$ and $I_{\mathrm{T}}$ is the total CR intensity (background plus nearby contribution) at that energy. The anisotropy $\delta_{\mathrm{m}}$ from the dominant single source under the diffusion approximation is given by (\citealt{Mao1972}),
\begin{eqnarray}
    \delta_{\mathrm{m}}=\frac{3D_\mathrm{L}}{c}\frac{|\nabla N_{\mathrm{m}}|}{N_{\mathrm{m}}}
\end{eqnarray}
where $N_{\mathrm{m}}$ is given by Equation (\ref{eq:nearby_solution}) for the dominant source with distance $r_{\mathrm{m}}$ and age $t_{\mathrm{m}}$, and $D_{\mathrm{L}}$ is the diffusion coefficient in the local region. For the proton spectrum shown in Figure \ref{fig:proton}, we get $\delta \sim (1.1-6)\times 10^{-3}$ over the energy range of ($1-100$) TeV. These values are approximately a factor of $(2-6)$ higher than the measured dipole anisotropy, which is $\sim (0.5-1)\times 10^{-3}$ in the same energy range (\citealp{Abeysekara_2019} and references therein). Similar discrepancies were also found in earlier studies (e.g., \citealt{Erlykin2006, Blasi_aniso, Thoudam2012b}). In our model, this discrepancy can be related to the poorly known diffusion coefficient in the local region and uncertainties in the age and distance parameters of the nearby SNRs. Cosmic rays probe large distances in the Galaxy, and therefore the diffusion coefficient determined from the boron-to-carbon data (available only up to a few TeVs/n) mainly represents the average value in the Galaxy (\citealp{Taillet2003, Thoudam2008}). This value can be different from that in the local region given that our Sun is located inside the Local Bubble, a hot cavity with an estimated size of $\sim\,0.5$~kpc (\citealp{Frisch2006}). Therefore, understanding the local magnetic field structure and its effects on the propagation of cosmic rays can be crucial for the anisotropy study. Indeed, the small-scale structures observed in anisotropy data at few TeVs are often attributed to the local magnetic field configuration (which is not so well understood) and to the nature of particle scattering in the field, as discussed in \cite{Drury2008}, \cite{Giacinti2012} and \cite{Ahlers2014}. Moreover, the effect of the heliosphere may be important. At energies below $\sim\,10$~TeV, cosmic rays have a maximum gyroradius of approximately $800$~AU\footnote{$1$~astronomical unit $\approx 4.85\times 10^{-9}$~kpc.} in a $3$~$\mu$G magnetic field, and therefore their propagation en route to the Earth can be further affected by the heliosphere which extends up to about a few thousands of AU (\citealp{Desiati2013, Pogorelov2016}).

%In addition, the discrepancy can arise from global effects such as uncertainties in the source distribution of cosmic rays in the Galaxy and the nature of cosmic-ray propagation through the Galactic magnetic field. For example, cosmic rays are expected to diffuse faster along the magnetic field lines than in the perpendicular direction (\citealp{kumar2014}).

%The calculated anisotropy for proton for Model 1 is $\delta = (0.85-4.2)\times10^{-3}$, and for Model 2 is $\delta= (0.78-3.8)\times 10^{-3}$, which are slightly larger than that of Model 1.

\section{Conclusion}
\label{section:conclude}
We have explored in detail the contribution of nearby SNRs to the observed flux of CRs at the Earth. Based on the results obtained in this work, we conclude that the spectral bump at TeV energies in the proton and helium spectra, recently observed by the CALET and the DAMPE  experiments, is most likely an effect of the contribution of nearby SNRs, in particular, the Vela supernova remnant. The contribution of the nearby SNRs is also found to be consistent with the observed spectra of heavy elements from carbon to iron, and also with the all-particle spectrum up to energies beyond the knee when combined with a background CR component originating from the distant sources.

\section*{acknowledgments}
SB acknowledges the Prime Minister's Research Fellowship (PMRF) and Govt. of India for financial support. S.T. acknowledges funding from the ADEK (AARE19-224) grant, and the Khalifa University ESIG-2023-008 and RIG-S-2023-070 grants. %PS acknowledges a Swarnajayanti Fellowship (DST/SJF/PSA-03/2016-17) and a National Supercomputing Mission (NSM) grant from the Department of Science and Technology, India.




\bibliography{sample631}{}
\bibliographystyle{aasjournal}



\end{document}